# A tropical geometry for bounded biochemical state spaces

James N. Cobley

The University of Dundee, Dundee, Scotland, UK.

**Correspondence**: (jcobley001@dundee.ac.uk) or (j_cobley@yahoo.com)

**ORCID**: https://orcid.org/0000-0001-5505-7424

**Abstract**
Many biochemical measurements define state spaces that are bounded, absorbing, and physically irreversible, yet are routinely analysed using linear and Euclidean frameworks that assume global invertibility, symmetry, and translation invariance. This mismatch can irretrievably obscure biological structure, independent of data quality, scale, or preprocessing. This work formalises the structure of bounded biochemical state spaces using cysteine redox regulation as a representative example and identify the minimal algebraic properties required for categorically correct representation. Hard boundaries, absorbing states, and irreversible ensemble dynamics render linear algebra incompatible with these objects. This work demonstrates that tropical algebra provides a natural realisation of the required properties by replacing additive linear structure with order-based, piecewise-linear operations that encode dominance, saturation, and path dependence without contradiction. By making non-invertibility and absorption explicit rather than implicit, this framework resolves a fundamental algebraic mismatch and establishes a principled foundation for the representation and analysis of bounded biochemical data.

**Introduction**
Categorically correct data representation and analysis require matching the algebra to the object of study in biology[1]. If the algebra is categorically mismatched, then normalisation, transformation, or modelling cannot recover the lost structure[2]. This loss arises from algebraic incompatibility rather than scale or noise. As a consequence, mismatched algebraic representations can produce concomitant losses of biological meaning, hampering discovery-driven analysis[3].

The regulation of biological processes by protein post-translational modifications (PTMs) is a core object of study in biology[4]. As a representative example of this paradigm, protein structure and function can be post-translationally regulated by chemically reversible cysteine oxidation events[5–9], which arise from metabolically coupled processes (e.g., reactive oxygen species production[10]). These events exert pleiotropic effects across the proteome[5–9], motivating the acquisition of expansive redox proteomic datasets[11–15] that require principled algebraic representation and analysis.

Like other PTMs, cysteine oxidation defines a mathematically bounded biochemical state space spanning a [0,1] interval[16]. While a single cysteine can be only be reduced (0) or oxidised (1), an ensemble of cysteines can take fractional values across the [0,1] interval[17]. These hard boundaries are physically immutable[18]—a cysteine proteoform molecule or population thereof cannot be more reduced than 0 or more oxidised than 1. The 0 boundary is an absorbing biological state as most cysteines are fully reduced[19–23]. In such a space, even an identical distance (e.g., 0.1) shift can encode a different meaning depending on the starting state (e.g., 0 to 0.1 vs. 0.5 to 0.6). While such shifts are chemically reversible at the molecular level[24], they are generically physically non-invertible at the ensemble level due to entropy[25].

Biochemical state spaces are routinely analysed using linear and Euclidean frameworks that assume global invertibility, symmetry, and translation invariance[26]. Common operations, such as normalisation, averaging, and dimensionality reduction, preserve these assumptions by construction[27], even when applied to bounded data[28]. However, these assumptions are categorically violated by hard boundaries, absorbing states, and generically physically irreversible dynamics[29–31]. Equal-magnitude changes do not encode equal biological meaning across the space, trajectories cannot be inverted, and intermediate states constrain the future evolution of the system[18]. Hence, linear algebra is categorically mismatched to bounded PTM biochemical state spaces.

To address this categorical algebraic mismatch, this work introduces a tropical algebra of bounded biochemical state spaces. Tropical algebra provides a category match by replacing additive linear structure with order-based, piecewise-linear operations that naturally encode dominance, saturation, and path-dependence. In this setting, boundaries are intrinsic features of the geometry and irreversible transitions are represented without contradiction. By resolving the categorical algebraic mismatch, this work establishes a foundation for the representation and analysis of bounded biochemical data more broadly.

**Results**
This theoretical work formalises the structure of bounded biochemical state spaces and identifies the minimal properties required for categorically correct representation. Rather than introducing a new analytical method, this work establishes that these properties impose algebraic constraints that are incompatible with linear and Euclidean representations, independent of normalization or pre-processing. This work formalises these constraints and provides intuitive geometric interpretations that motivate the algebraic resolution introduced subsequently. Formal definitions, propositions, and proofs are provided in the Supplemental Notes.

**Structure of bounded biochemical state spaces**
Many biochemical measurements are naturally described as state variables constrained to a finite interval (**Figure 1**). As a representative case, cysteine redox states are commonly quantified as an occupancy bounded between fully reduced (0) and fully oxidised (1) limits[16]. These bounds are physically immutable constraints: a cysteine molecule or population cannot be more reduced than 0 or more oxidised than 1[18]. These scale-invariant bounds apply to both the protein-level object—the cysteine proteoform[32–34]—and its component parts—each individual cysteine site. Resultantly, the underlying state space is closed and bounded, and algebraic operations that assume unbounded translation are immediately incompatible.

The boundaries of this space act as absorbing states[20]. A large fraction of cysteines reside at or near the fully reduced boundary[11], and many distinct biochemical trajectories collapse onto the same observed state. Once this collapse occurs, information about prior history is irretrievably lost at the level of observation—the act of measurement. Consequently, the mapping from biochemical trajectories to measured states is generically many-to-one, rendering the state space non-invertible. For example, for a cysteine proteoform with three cysteines considered in redox binary, both the 000 state could arise from any one of the 100, 010, and 001 biochemical states (3:1 mapping)[18].

In such bounded spaces, equal-magnitude changes do not encode equal biological meaning. For example, a shift from 0.0 to 0.1 represents the emergence of oxidation from an absorbing boundary, whereas a shift from 0.5 to 0.6 reflects redistribution within an already oxidised ensemble. Conceptually, this represents the difference between emergent "switch" and "toggle" behaviour, respectively[28]. Although numerically identical, these changes differ fundamentally in their physical interpretation and in the constraints they impose on subsequent state evolution[18].

While individual redox reactions may be chemically reversible[24], ensemble-level redox dynamics are generically physically irreversible due to entropy production[29–31] (e.g., via heat dissipation). Transitions that reach or approach absorbing boundaries reduce the accessible state space for future evolution in a given direction (e.g., towards the fully reduced state), introducing path dependence into the system[18]. As a result, the sequence in which biochemical transitions occur matters: intermediate states constrain subsequent possibilities even when the net chemical change is nominally reversible.

Together, hard boundaries, absorbing states, and physical irreversibility define the essential structure of bounded biochemical state spaces[35]. These properties are intrinsic to the object and must be respected by any algebraic representation intended to analyse such data. Formal definitions and proofs of the consequences of these properties are provided (**Supplemental Note 1**).

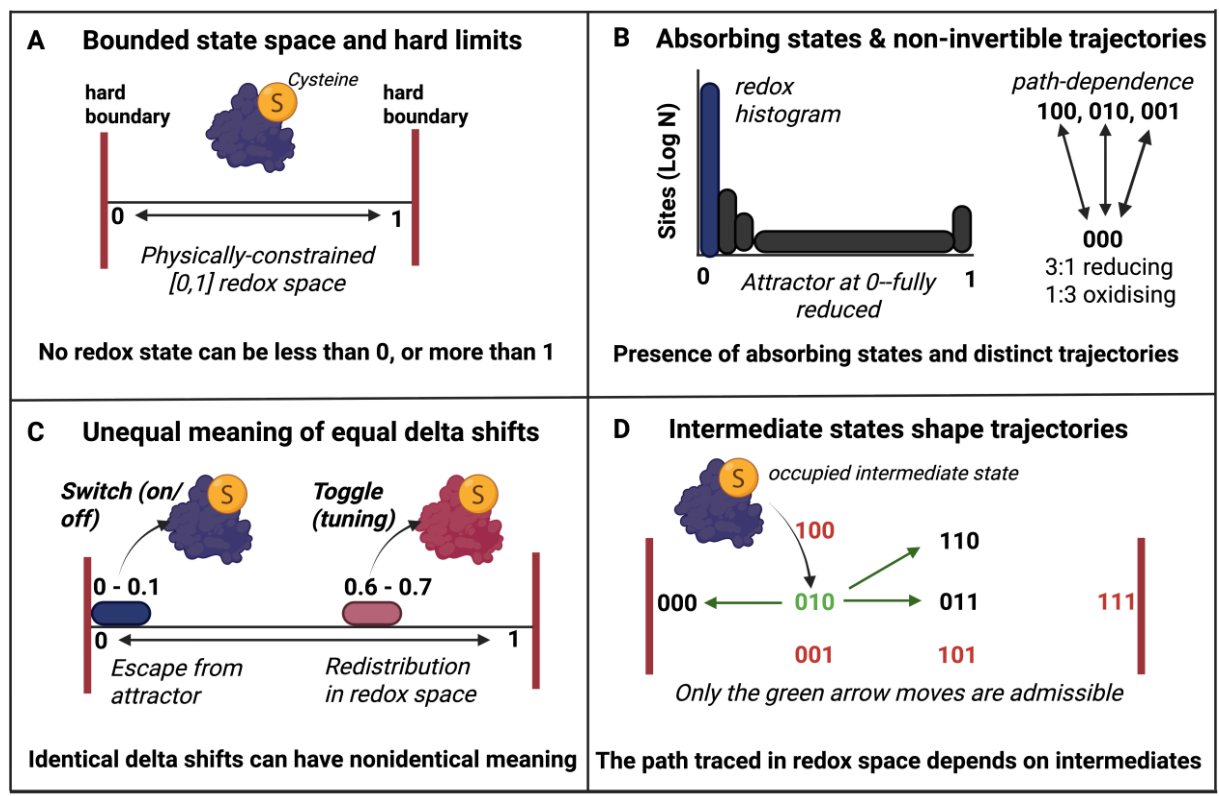

**Figure 1. Structural properties of bounded biochemical state spaces illustrated by cysteine redox.**
(A) Cysteine redox states are constrained to a physically immutable, closed interval $[0, 1]$, corresponding to fully reduced (0) and fully oxidised (1) ensembles. No biochemical or observational state can exist outside these hard boundaries.
(B) The bounded state space contains absorbing states, illustrated by the strong empirical bias toward the fully reduced boundary. Distinct biochemical trajectories collapse onto common observed states, producing a many-to-one mapping and rendering the space non-invertible.
(C) Equal-magnitude changes in redox occupancy encode unequal biological meaning depending on position within the bounded space. Small shifts near the reduced boundary correspond to switch-like activation (escape from an attractor), whereas identical shifts within the interior reflect tuning or redistribution within an already oxidised ensemble.
(D) Intermediate states constrain admissible trajectories in discrete proteoform space. Only transitions consistent with occupied intermediates are allowed, introducing path dependence even when individual redox reactions are chemically reversible.

**Algebraic incompatibility of linear representations**
Linear and Euclidean representations assume global invertibility, symmetry, and translation invariance of the underlying state space[26]. These assumptions are preserved under common operations such as normalization, averaging, and dimensionality reduction, which rescale or reorient data without altering its algebraic structure. As formally demonstrated (**Supplemental Note 2**), the properties of bounded biochemical state spaces are categorically incompatible with the assumptions of linear algebra.

**Minimal algebraic requirements for bounded state spaces**

Any algebraic framework suitable for representing bounded biochemical state spaces must satisfy a minimal set of requirements imposed by the object itself. The algebra must:

(i) respect hard bounds without relying on unconstrained translation;
(ii) accommodate absorbing states and many-to-one mappings without assuming global invertibility;
(iii) represent irreversible, path-dependent transitions without assuming symmetry or commutativity;
(iv) remain well-defined on discrete or combinatorial state spaces[36], such as proteoform ensembles[37–39].

These requirements follow directly from the structural properties established above and therefore constrain any admissible algebraic representation.

While this work does not claim it is the only representation, tropical algebra provides a natural realisation of these requirements by replacing additive linear structure with order-based, piecewise-linear operations. In tropical geometry, bounds are intrinsic rather than imposed, absorbing states arise naturally through dominance, and irreversible transitions are represented without contradiction. Tropical operations remain well-defined on discrete state spaces, allowing direct representation of proteoform-level objects without invoking infinitesimal neighbourhoods or smooth flows.

**Tropical representation of bounded biochemical state spaces**

To make the algebraic representation explicit[40–42], this work encodes cysteine redox states as bounded occupancies $x \in [0,1]$, where $x = 0$ denotes a fully reduced ensemble and $x = 1$ denotes a fully oxidised ensemble. Changes in redox state therefore correspond to ordered transitions within a closed interval—a bounded tropical semiring—rather than additive displacements on an unbounded axis (**Figure 2**).

In a tropical representation, linear addition is replaced by order-based operations acting directly on a bounded biochemical observable. For a residue-level occupancy $x \in [0,1]$, tropical addition is defined as

$$a \oplus b = \max(a, b),$$

which encodes dominance and biological saturation. Sequential biochemical transitions are composed via

$$a \otimes b = \min(1, a + b),$$

ensuring closure on the bounded state space. Here, $x$ denotes the current biochemical state, while $b$ represents a residue-level redox increment; state evolution is defined by the tropical action of such increments on the current occupancy. Under this formulation, oxidation events increase occupancy toward the upper boundary, reduction events decrease occupancy toward the lower boundary, and operations that leave the state unchanged under $\oplus$ encode biological saturation rather than the absence of activity.

Non-invertibility arises naturally in this framework as a structural consequence of the bounded, absorbing state space. Multiple distinct biochemical trajectories may converge to the same boundary value, irreversibly collapsing information about prior history at the level of observation. Once this collapse occurs, no global inverse mapping can exist, even though individual redox reactions may be chemically reversible. Tropical geometry makes this loss of invertibility explicit rather than concealing it behind assumptions of smoothness or reversibility.

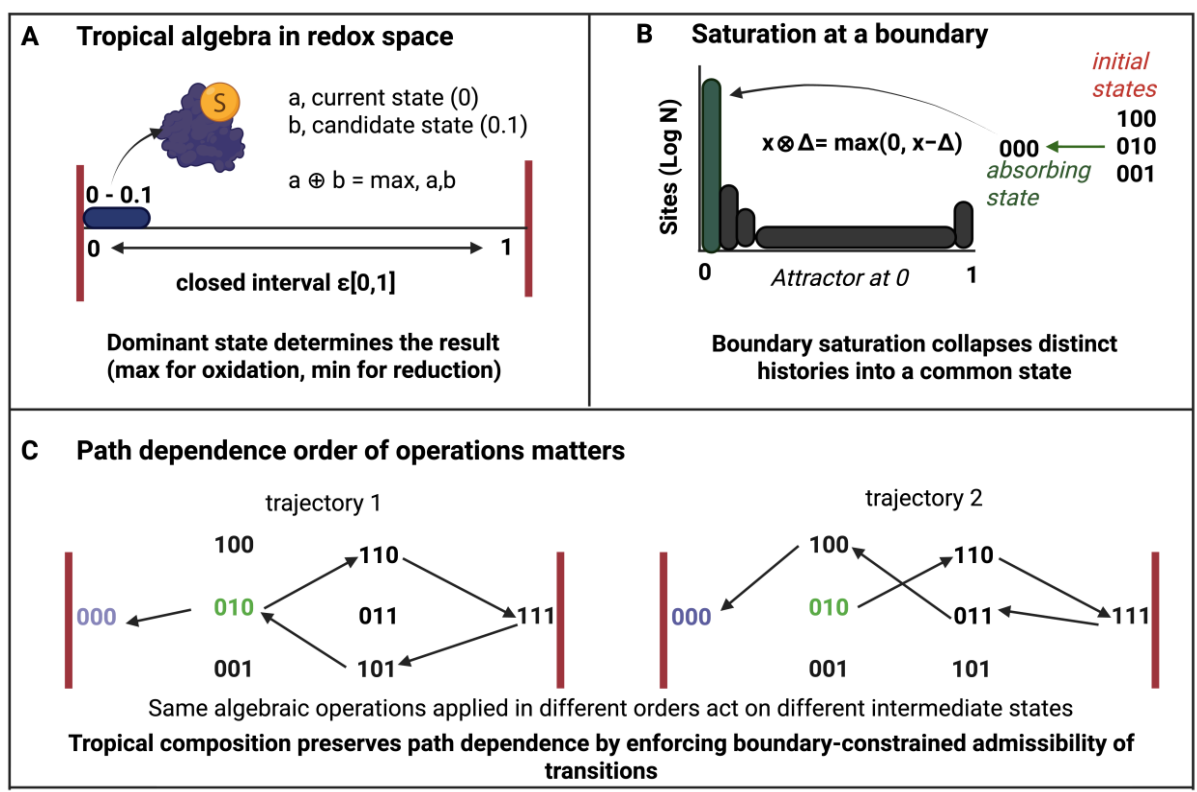

**Figure 2 | Tropical algebra provides a categorically correct representation of bounded redox state evolution.**
**(A)** Tropical addition acting on residue-level redox occupancies $x \in [0,1]$ selects the dominant state via order-based comparison $(a \oplus b = \max(a,b))$, enforcing a closed, physically constrained interval. Oxidation and reduction are represented without invoking unbounded translation.
**(B)** Boundary saturation at the fully reduced state $(x = 0)$ acts as an absorbing state. Distinct initial proteoform configurations collapse onto a common observed state under boundary-constrained composition, rendering the mapping from trajectories to observations many-to-one and non-invertible.
**(C)** Path dependence emerges from boundary-constrained admissibility of transitions. The same algebraic operations applied in different orders act on different intermediate states, yielding distinct trajectories even when the net operations are identical. Tropical composition preserves this path dependence by enforcing state-dependent constraints on allowable transitions.

It is important to distinguish between algebraic operations and state evolution under their composition. Individual tropical operations are typically commutative (for example, $a \oplus b = b \oplus a$). However, the induced evolution of biochemical states is not commutative in bounded, irreversible spaces. The order in which redox events—and more generically PTM

events—occur matters because intermediate states constrain subsequent possibilities. An oxidation followed by reduction need not yield the same outcome as reduction followed by oxidation if one sequence encounters a boundary or dissipative loss before the other.

A similar distinction applies to associativity. While tropical operations themselves are associative, the evolution of biochemical states under grouped sequences of transitions becomes path-dependent due to boundary constraints. Grouping transitions differently can lead to distinct intermediate states, and these intermediate states affect the admissible set of future transitions. As a result, biochemical state evolution is inherently path-dependent, even when individual transformations are well defined.

Hard boundaries therefore play an active and explicit role in tropical representations. Once a state reaches a boundary, further operations in the same direction leave the state unchanged, reflecting biological saturation rather than mathematical stagnation. In this sense, boundaries act as absorbing states: they terminate certain classes of trajectories and collapse distinct histories into a common endpoint, analogous to extremal regions in tropical geometry where dominance becomes degenerate.

As a concrete illustration, consider a single cysteine residue with initial occupancy $x = 0.92$. Under a linear representation, an oxidation increment of $+0.1$ would formally yield $x = 1.02$, a state that is biologically and observationally impossible. In the tropical representation, the same transition is composed as

$$x \otimes 0.1 = \min(1, 0.92 + 0.1) = 1,$$

explicitly encoding saturation at the hard boundary. Any subsequent oxidation leaves the state unchanged under $\oplus$, while reduction transitions act relative to this saturated state. Distinct oxidation histories that reach the boundary therefore collapse onto the same observed value, making the loss of invertibility explicit rather than implicit.

Formal definitions of the bounded tropical state space, proofs of non-invertibility and path dependence, and extensions to discrete proteoform ensembles are provided (**Supplemental Notes 3-4**).

**Discussion**

Biochemical measurements often define state spaces that are bounded, absorbing, and physically irreversible. This work formalised these properties, demonstrating that they are intrinsic to many PTMs, including cysteine redox regulation. These structural features impose algebraic constraints that are categorically incompatible with linear and Euclidean representations, independent of normalisation, transformation, or modelling choices.

This incompatibility does not arise from noise, scale, or insufficient data, but from a mismatch between the algebra commonly applied and the object being represented. In bounded biochemical state spaces, trajectories are generically non-invertible, equal-magnitude changes do not encode equal biological meaning, and intermediate states constrain future evolution. Linear frameworks, which assume global invertibility, symmetry, and translation invariance, therefore conflate distinct biochemical histories and obscure the geometry of the underlying state space.

This work identified a minimal set of algebraic requirements imposed by such bounded objects and showed that tropical algebra provides a natural realisation of these requirements. By replacing additive linear structure with order-based, piecewise-linear operations, tropical representations encode dominance, saturation, and path dependence without contradiction. Importantly, tropical algebra does not impose irreversibility or absorption artificially, but makes these properties explicit where they already exist in the biology.

While cysteine redox regulation served as a concrete motivating example, the arguments presented here are not specific to redox biology. Many biochemical and molecular measurements define bounded state spaces with absorbing boundaries and irreversible ensemble dynamics. This work, therefore, applies broadly to the representation of bounded biological data, wherever categorical mismatches between object and algebra arise.

This work does not claim that tropical algebra is the only possible resolution of such mismatches, nor that all biological systems require tropical representations. Rather, it establishes a general principle: categorically correct analysis requires that the algebra respect the structural properties of the object under study. When this condition is violated, no amount of post hoc transformation can recover the lost structure. Ensuring this match is a prerequisite for meaningful representation, interpretation, and discovery in biological data analysis. It is envisaged that tropical algebraic representations and analysis of PTM data will yield novel insights.

**Acknowledgements.**
The author thanks Prof. Angus I. Lamond (the University of Dundee) and all of the members of the Lamond lab for constructive scientific discussions. The support of an MCR grant (MR/Y013891/1) is gratefully acknowledged. Figure 1 and 2 were created using BioRender and exported at 600 DPI PNG files with a publication license.

**Conflict of interest**
The author declares that there are no conflicts of interest.

**Declaration of generative AI and AI-assisted technologies in the writing process**
During the preparation of this work, the author used ChatGPT (OpenAI) as an AI-assisted tool to support language editing, structural refinement, and clarification of conceptual arguments. All scientific content, theoretical development, interpretations, and conclusions were conceived by the author. The author reviewed, edited, and verified all text generated with AI assistance and takes full responsibility for the content of the published article.

**Supplemental Notes**
**Supplemental Note 1 — Formal structure of bounded biochemical state spaces**
**S1.1. State space and observables**
Let a biochemical modification state be represented by a bounded occupancy
$$x \in \mathcal{X} := [0,1],$$

where $x = 0$ denotes fully unmodified (e.g., fully reduced) and $x = 1$ denotes fully modified (e.g., fully oxidised). For proteoform-level objects comprising $r$ sites, we consider the product space
$$\mathcal{X}^{(r)} := [0,1]^r.$$

In many PTM contexts, the measured quantity is an ensemble occupancy (fraction modified), and therefore a coarse-grained observable. We model this via a measurement map
$$M: \Omega \to \mathcal{X},$$

from a (typically unobserved) space of micro-trajectories $\Omega$ to a bounded observable $x$.

**S1.2. Hard bounds and closure**
The state space $\mathcal{X}$ is closed and bounded. Any operation interpreted as a "translation" $x \mapsto x + \delta$ with $\delta \neq 0$ is not closed on $\mathcal{X}$ in general, since $x + \delta \notin [0,1]$ whenever $x$ is sufficiently close to a boundary.

A natural way to represent finite changes while preserving closure is by *saturated* (clipped) maps. For example, define oxidation and reduction updates of magnitude $a, b \geq 0$ by
$$\mathrm{Ox}_a(x) := \min(1, x + a), \mathrm{Red}_b(x) := \max(0, x - b).$$

Then $\mathrm{Ox}_a, \mathrm{Red}_b: \mathcal{X} \to \mathcal{X}$ are well-defined endomorphisms.

**S1.3. Absorbing boundaries**
**Definition (absorbing boundary).** A boundary point $x^\star \in \{0,1\}$ is absorbing for an update family $\{F_\theta\}$ if for all admissible $\theta$ in a direction,
$$F_\theta(x^\star) = x^\star.$$

**Proposition S1 (absorption under saturation).** For all $a \geq 0$ and $b \geq 0$,
$$\mathrm{Ox}_a(1) = 1, \mathrm{Red}_b(0) = 0.$$

Hence 1 is absorbing for oxidation updates and 0 is absorbing for reduction updates.
*Proof.* Immediate from the definitions: $\min(1, 1 + a) = 1$ and $\max(0, 0 - b) = 0$. □

**S1.4. Many-to-one mapping and non-invertibility**
**Proposition S2 (many-to-one collapse).** For any $a > 0$,
$$\mathrm{Ox}_a(x) = 1 \text{ for all } x \in [1 - a, 1].$$

Similarly, for any $b > 0$,
$$\mathrm{Red}_b(x) = 0 \text{ for all } x \in [0, b].$$

*Proof.* If $x \geq 1 - a$, then $x + a \geq 1$ so $\min(1, x + a) = 1$. The reduction case is analogous. □

**Corollary S2.1 (non-invertibility).** For any $a > 0$, $Ox_a$ is not injective (hence not invertible) on $\mathcal{X}$. Likewise $Red_b$ for any $b > 0$.

This formalises the claim that distinct biochemical histories can collapse onto the same observed boundary state.

### S1.5. Equal-magnitude changes encode unequal meaning (position dependence)

For a fixed increment $\delta > 0$, the *available capacity* to move upward without saturation at state $x$ is $c_+(x) := 1 - x$; downward is $c_-(x) := x$. A change of magnitude $\delta$ is therefore interpreted relative to state-dependent capacity.

**Proposition S3 (position dependence).** For $\delta > 0$, the "same" increment $\delta$ produces a saturated outcome iff $x \geq 1 - \delta$. Hence, the effect of a $\delta$-update depends on $x$ through $c_+(x)$.

This captures formally why a shift $0 \to 0.1$ and $0.5 \to 0.6$ are not equivalent: only the former crosses the absorbing neighbourhood where collapse is common and capacity is minimal.

### S1.6. Proteoform ensembles induce intrinsic many-to-one structure

Let $r$ binary sites define a proteoform state space $\{0, 1\}^r$. A proteoform ensemble is a distribution $p$ on $\{0, 1\}^r$, i.e. $p \in \Delta_{2^r}$ (the probability simplex). The measured site occupancies are the marginals

$$m_i = \mathbb{E}_p[s_i], i = 1, \ldots, r,$$

defining a mapping

$$\pi: \Delta_{2^r} \to [0,1]^r, \pi(p) = (m_1, \ldots, m_r).$$

**Proposition S4 (non-invertibility of marginalisation).** For $r \geq 2$, the map $\pi$ is generically many-to-one: distinct proteoform distributions can yield identical marginal occupancies.

*Proof (constructive).* For $r = 2$, let $p^{(1)}$ place mass $(1/2, 0, 0, 1/2)$ on $(00, 01, 10, 11)$ and let $p^{(2)}$ place mass $(0, 1/2, 1/2, 0)$. Both yield marginals $m_1 = m_2 = 1/2$ but $p^{(1)} \neq p^{(2)}$. The general case follows by embedding this 2-site construction within $r$ sites. □

This shows that many-to-one structure is present even before considering dynamics: it is intrinsic to coarse-grained measurement of proteoform ensembles.

## Supplemental Note 2 — Incompatibility of bounded, absorbing state spaces with linear/Euclidean representations

This note formalises the sense in which linear and Euclidean frameworks assume properties (global translation, invertibility, symmetry) that are violated by bounded, absorbing state spaces.

### S2.1. Linear structure implies global translation and inverses

A real vector space $(V, +)$ is an abelian group under addition: for every $v \in V$, an additive inverse $-v$ exists with $v + (-v) = 0$, and translation maps $T_u(v) = v + u$ are bijections of $V$.

By contrast, bounded biochemical state spaces $\mathcal{X} = [0,1]$ do not admit a globally closed translation operation $x \mapsto x + \delta$ for arbitrary $\delta \neq 0$. Any attempt to enforce closure by saturation introduces many-to-one collapse near boundaries (Proposition S2), which destroys bijectivity and therefore invertibility.

### S2.2. No globally invertible "translation-like" dynamics on $[0, 1]$ with absorption

**Proposition S5 (absorption implies non-bijectivity).** Let $F: \mathcal{X} \to \mathcal{X}$ be any map such that there exists an interval $I \subseteq \mathcal{X}$ with $F(x) = x^\star$ for all $x \in I$ (collapse to an absorbing state). Then $F$ is not injective and has no inverse on $\mathcal{X}$.

*Proof.* If $|I| > 1$, pick $x \neq y \in I$; then $F(x) = F(y) = x^\star$, so $F$ is not injective. □

Because absorption/collapse is an intrinsic feature of the object (Supplemental Note 1), any representation that assumes invertible dynamics is categorically mismatched.

### S2.3. Euclidean distance is translation-invariant; bounded saturation is not

In Euclidean spaces, distances satisfy translation invariance:
$$\| (v + u) - (w + u) \| = \| v - w \| \quad \forall u, v, w.$$

In bounded state spaces with saturation, the effect of applying the "same" increment depends on position (Proposition S3), so translation-invariance fails at the level of dynamics. Consequently, Euclidean interpretations of "equal distances imply equal meaning" are not valid for state evolution on $[0, 1]$ in the presence of absorbing boundaries.

## Supplemental Note 3 — A bounded tropical semiring for biochemical occupancies

### S3.1. Definition (bounded tropical operations)

Let $\mathbb{T}_{[0,1]} := ([0,1], \oplus, \otimes)$ with operations
$$a \oplus b := \max(a, b),$$
$$a \otimes b := \min(1, a + b).$$

(Here $\otimes$ is *capped addition*, ensuring closure on $[0, 1]$.)

### S3.2. Algebraic properties

**Proposition S6 (closure).** For all $a, b \in [0,1]$, $a \oplus b \in [0,1]$ and $a \otimes b \in [0,1]$.

*Proof.* $\max(a, b) \in [0,1]$. Also $a + b \in [0,2]$ so $\min(1, a + b) \in [0,1]$. □

**Proposition S7 (commutativity and associativity).** $\oplus$ and $\otimes$ are commutative and associative on $[0, 1]$.

*Proof.* Commutativity is immediate. Associativity of $\oplus$ follows from associativity of max. For $\otimes$,
$$(a \otimes b) \otimes c = \min(1, \min(1, a+b) + c) = \min(1, a+b+c) = a \otimes (b \otimes c).$$
□

**Proposition S8 (identities and absorption).** The element 0 is the identity for both operations:
$$a \oplus 0 = a, \quad a \otimes 0 = a,$$
and 1 is absorbing for $\otimes$:
$$a \otimes 1 = 1.$$

*Proof.* Immediate from definitions. □

This mirrors biochemical structure: 1 acts as saturation (absorbing under further oxidation-like composition).

**Proposition S9 (idempotency and non-existence of additive inverses).** $\oplus$ is idempotent: $a \oplus a = a$. Consequently, $([0,1], \oplus)$ cannot be a group and admits no global inverses compatible with $\oplus$.

*Proof.* $a \oplus a = \max(a, a) = a$. If $\oplus$ had inverses, idempotency would force $a = 0$ for all $a$, a contradiction. □

This provides a clean algebraic statement of non-invertibility without appealing to dynamics.

**Supplemental Note 4 — Order dependence and non-commutativity of biochemical transition sequences**

The main text distinguishes commutativity of tropical operations from order dependence of *state transition sequences*. We formalise that distinction here using oxidation/reduction operators on $[0, 1]$.

**S4.1. State transition operators**

Define oxidation and reduction operators as in S1:
$$\mathrm{Ox}_a(x) = \min(1, x + a), \quad \mathrm{Red}_b(x) = \max(0, x - b).$$

These maps are monotone and closed on $[0, 1]$.

**S4.2. Non-commutativity of mixed transition sequences**

**Proposition S10 (order dependence).** There exist $a, b > 0$ and $x \in [0,1]$ such that
$$(\mathrm{Ox}_a \circ \mathrm{Red}_b)(x) \neq (\mathrm{Red}_b \circ \mathrm{Ox}_a)(x).$$

*Proof (constructive).* Choose $x = 1 - \varepsilon$ with $0 < \varepsilon < a$, and choose $b > a - \varepsilon$. Then
$$(\mathrm{Ox}_a \circ \mathrm{Red}_b)(x) = \mathrm{Ox}_a(1 - \varepsilon - b) = \min(1, 1 - \varepsilon - b + a),$$

while
$$(\mathrm{Red}_b \circ \mathrm{Ox}_a)(x) = \mathrm{Red}_b(1) = \max(0, 1 - b) = 1 - b.$$

With the above choices, these quantities differ (one path saturates before reduction, the other reduces after saturation). □

This shows that although $\oplus$ is commutative, *mixed biochemical update sequences* need not commute because boundaries change the effect of subsequent operations.

**S4.3. Path dependence via collapse**

**Proposition S11 (history erasure near boundaries).** For any $a > 0$, the set $[1 - a, 1]$ is collapsed by $\mathrm{Ox}_a$ to the single value 1. Therefore, trajectories entering this region become indistinguishable under subsequent observations.

This formalises "irretrievable loss of history" at the observation level and explains why ensemble dynamics are physically non-invertible.